\documentclass[12pt,3p,authoryear]{elsarticle}
\usepackage[T1]{fontenc}
\usepackage{amsmath,amssymb,amsfonts}
\usepackage{natbib}
\usepackage{graphicx}
\usepackage{epstopdf}
\usepackage[hidelinks,breaklinks=true]{hyperref}
\usepackage{booktabs}
\usepackage{rotating}
\usepackage{times}
\usepackage{xspace}
\usepackage{dcolumn}
\newcolumntype{.}[1]{D{.}{.}{#1}}
\newcolumntype{,}[1]{D{,}{,}{#1}}
\usepackage{setspace}
\usepackage{color}
\usepackage{subdepth}
\usepackage{needspace}   
\usepackage{float}
\usepackage[font={small,singlespacing},labelfont={bf,small},skip=0.2cm]{caption}
\floatstyle{plaintop}
\restylefloat{table}
\restylefloat{figure}			
\journal{Finance Research Letters}
\graphicspath{{./img/}}
\usepackage{multirow}
\RequirePackage{ae,fancyvrb}
\newcommand{\pkg}[1]{\textbf{#1}}

\newcommand{\proglang}[1]{\textsf{#1}}
\usepackage{upquote}
\DefineVerbatimEnvironment{CodeInput}{Verbatim}{fontshape=sl}
\DeclareMathAlphabet\mathbfcal{OMS}{cmsy}{b}{n}

\usepackage[font={normal}]{subfig}

\usepackage{color}

\newcommand{\ie}{\emph{i.e.}\xspace}
\newcommand{\eg}{\emph{e.g.}\xspace}

\newcommand{\insertfloat}[1]{%
\begin{center}
[Insert~#1 about here.]%
\end{center}%
}

\begin{document}
\begin{frontmatter}
\title{Factor Exposure Heterogeneity in Green and Brown Stocks\tnoteref{label1}}
\tnotetext[label1]{We are grateful to IVADO, the Natural Sciences and Engineering Research Council of Canada (grant RGPIN-2022-03767), and the Swiss National Science Foundation (grant \#179281) for their financial support.}
\author[hec]{David Ardia\corref{cor1}}
\ead{david.ardia@hec.ca}
\cortext[cor1]{Corresponding author. HEC Montréal, 3000 Chemin de la Côte-Sainte-Catherine, Montreal, QC H3T 2A7. Phone: +1 514 340 6103.}
\ead{keven.bluteau@usherbrooke.ca}
\author[sherbrooke]{Keven Bluteau}
\ead{keven.bluteau@usherbrooke.ca}
\author[hec]{Gabriel Lortie-Cloutier}
\ead{gabriel.lortie-cloutier@hec.ca}
\author[hec]{Thien Duy Tran}
\ead{thien-duy.tran@hec.ca}
\address[hec]{GERAD \& Department of Decision Sciences, HEC Montréal, Montréal, Canada}
\address[sherbrooke]{Department of Finance, Université de Sherbrooke, Canada\\[1cm]
\large Published in Finance Research Letters\\
\url{https://doi.org/10.1016/j.frl.2023.103900}\\
CC-BY-NC-ND License}

\begin{abstract}
Using the peer-exposure ratio, we explore the factor exposure heterogeneity in green and brown stocks. By looking at peer groups of S\&P 500 index firms over 2014-2020 based on their greenhouse gas emission levels, we find that, on average, green stocks exhibit less factor exposure heterogeneity than brown stocks for most of the traditional equity factors but the value factor. Hence, investment managers shifting their investments from brown stocks to green stocks have less room to differentiate themselves regarding their factor exposures. Finally, we find that factor exposure heterogeneity has increased for green stocks compared to earlier periods.
\end{abstract}
\end{frontmatter}

\doublespacing

\newpage
\section{Introduction}

\noindent
Environmental, social, and governance (ESG) dimensions have become key for investors and asset managers in recent years \citep{GutscheZiegler2019}. On the environmental side, asset managers are more and more required to limit or avoid investing into brown stocks, that is, stocks of firms that are generating, directly or indirectly, high greenhouse gas (GHG) emission levels. However, the integration of ESG considerations into investment decision-making raises several challenges. Indeed, limiting the investment universe can result in (i) reduced alpha opportunities and/or (ii) limited exposures to traditional factors \citep{DiazEtAl2021}. 
As the average return investment managers deliver to investors is the sum of the alpha component (\ie, unconditional performance) and the beta component (\ie, performance resulting from being exposed to factors), measuring alpha and beta heterogeneity in specific green/brown universes is of practical relevance to determine if managers can still distinguish themselves from their peers when constrained to a given investment universe ---and so, typically, when shifting from brown to green stocks.
 
In previous work, we looked at the alpha component heterogeneity \citep{ArdiaEtAl2022}. Specifically, we investigated to what extent investment managers can differentiate themselves in terms of future alpha performance when focusing on green or brown stocks' universes. Focusing on S\&P 500 index firms over 2014--2020 and defining peer groups in terms of firms' GHG emission levels, we found that, on average, about 20\% of the stocks differentiate themselves from their peers in terms of alpha. Furthermore, we showed that the alpha heterogeneity has decreased over time, especially for green stocks, implying that it is more difficult for investment managers to deploy their skills when choosing among low-GHG intensity stocks in recent periods.

Looking at the alpha component only tells part of the performance story, as most investment managers have a clear factor exposures mandate. It is especially true for managers deploying smart beta/factor investing solutions, for which performance comes from the beta component they sell to their clients \citep{DimsonEtAl2017}. To our knowledge, a handful of studies look at the impact of ESG investment constraints on factor exposures. \citet{MelasEtAl2017} show that ESG scores positively correlate with the size, quality, and low volatility factors. \citet{BenderEtAl2017} find that highly related ESG companies have a positive value exposure and highlight that the factor exposure is not constant over time. \citet{MadhavanEtAl2021} find that funds with high environmental scores have increased quality and momentum factor exposures. Overall, these studies look at the level of exposure with respect to ESG ratings and/or the impact of ESG criteria on the firm performance. Our note aims to provide new and complementary insights by investigating not only the level but the heterogeneity in factor exposures achievable in green and brown stocks' universes. Our analyses will typically enable managers to determine for which factors they can deploy factor exposure tilting when moving their brown investments to green investments.

To explore the factor exposure heterogeneity, we rely on the peer-exposure ratio, an adaption of the peer-performance ratio proposed in \citet{ArdiaBoudt2018}. Instead of looking at the alpha differential between stocks within peer groups of green and brown stocks as in \citet{ArdiaEtAl2022}, we look at the beta differentials between stocks. Specifically, for each factor, we compute an equal-exposure ratio that measures the percentage of stocks in a given peer group that cannot be (significantly) differentiated in terms of their estimated factor's beta. 

Using a six-factor model based on the five factors of \citet{FamaFrench2015} and the momentum factor of \citet{Carhart1997}, we find that, on average, green stocks exhibit less factor exposure heterogeneity than brown stocks for most of the traditional equity factors but the value factor. Hence, investment managers shifting their investments from brown stocks to green stocks have less room to differentiate themselves regarding their factor exposures. Finally, compared to earlier periods, we find that factor exposure heterogeneities have increased in both universes. This contrasts with the significant decrease in alpha component heterogeneity observed for green stocks in \citet{ArdiaEtAl2022}.

\section{Data}
\label{sec:data}

\noindent
Our analysis focuses on  S\&P 500 companies between January 2014 and December 2020. Similar to \citet{ArdiaEtAl202x}, we define green (brown) firms as firms that create economic value while minimizing (not minimizing) damages that contribute to climate change. To classify firms, we use the greenhouse gas (GHG) emission intensity, which is CO2-equivalent emissions scaled by the firm's revenues. The data are extracted from Thomson/Refinitiv for the CO2-equivalent emissions data, Compustat for the firms' revenue, CRSP for the daily stock prices, and finally from Kenneth French's website\footnote{See \url{http://mba.tuck.dartmouth.edu/pages/faculty/ken.french/data_library.html}.} for the daily factor data. 

\section{Methodology}
\label{sec:methodology}

\noindent
To measure the heterogeneity in factor exposures investment managers face by selecting green or brown stocks for the 2009--2020 evaluation period, we adapt the methodology in \citet{ArdiaEtAl2022} as follows.

\paragraph{Step 1} In a given month, we form peer groups of brown and green stocks using the information available up to that month. We rely on the firms' latest GHG emissions to form the groups.\footnote{GHG data are usually released with a one-year delay.} Brown (green) stocks belong to the top 75th (bottom 25th) percentile of the GHG intensities.\footnote{Alternative thresholds were also tested and produced similar results. They are available from the authors upon request.} Each group contains the same number of peers.

\paragraph{Step 2} For each firm $i=1,\ldots,N$ within a peer group, we calculate the equal-exposure ratio $\pi_{i,k}^0$ for the $k=1,\ldots,K$ factors.  The ratio $\pi_{i,k}^0$ measures the percentage of peers stocks equally exposed to the factor $k$ as stock~$i$. This ratio is obtained similarly to the method presented in \citet[][Section 2]{ArdiaBoudt2018}, but adapted to beta calculations in the following way:

\paragraph{Step 2a} For each of the $N-1$ peer stocks $\mathcal{J}_i = \{j=1,\ldots,N; j \neq i\}$, we calculate the $p$-value of the null hypothesis of equal-exposure to factor $k$ over a forward-looking evaluation period between stock $i$ and stock $j \in \mathcal{J}_i$ using a pairwise test. We consider a $h$-month-ahead horizon of $T$ daily returns and estimate the beta differentials from a factor model:
\begin{equation}
r_{i,t}-r_{j,t}=\Delta \alpha_{i,j}+\sum_{k=1}^K \Delta \beta_{i,j,k} F_{k,t} + \varepsilon_{i,j,t} \,\, (j \in \mathcal{J}_i \,;\, t=1,\ldots,T) \,,
\end{equation}
where $r_{i,t}$ is the daily return of stock $i$ at time $t$, $F_{k,t}$ is the daily return of the $k$th factor at time $t$, and $\varepsilon_{i,j,t}$ is an error term. In our empirical application, we rely on a six-factor model using the five factors in \citet{FamaFrench2015} augmented with the momentum factor of \citet{Carhart1997}. Specifically, the six-factor model contains the market (MKT), the size (small-minus-big or SMB), the value (high-minus-low or HML), the profitability (robust-minus-weak or RMW), the investment (conservative-minus-aggressive or CMA), and the momentum (MOM) factors. This leads to a set of coefficients $\{\Delta \hat \beta_{i,j,k} , j \in \mathcal{J}_i\}$ for $k=1,\ldots,K$. Each coefficient is then standardized using the heteroscedastic and autocorrelation robust (HAC) standard error estimator of \citet{Andrews1991} and \citet{AndrewsMonahan1992}, leading to a studentized test statistic from which the $p$-value is obtained via the probability integral transform.

\paragraph{Step 2b} The distribution of the $p$-values obtained in \emph{Step 2a} is (asymptotically) a mixture of $p$-values that are uniformly distributed when the null hypothesis is true and $p$-values close to zero when the null hypothesis is false. Following \citet[Section 2.4]{ArdiaBoudt2018}, this allows us to set a data-driven cut-off point to estimate the proportion of equal-exposure $\pi_{i,k}^0$, which is robust to false discoveries.\footnote{All of the computations employed the \proglang{R} statistical computing language \citep{R} with the package \textbf{PeerPerformance} \citep{PeerPerformance}, which is freely available at \url{https://CRAN.R-project. org/package=PeerPerformance}.}

\paragraph{Step 3} Once the ratios are computed for the $N$ firms within a peer group, we aggregate them to $\pi_k^0 = \frac{1}{N} \sum_{i=1}^N \pi_{i,k}^0$ that measures the equal-exposure ratio for factor $k$ within a peer group. Thus, $1-\pi_k^0$ can be used to assess the extent to which stocks have a (significant) different exposure to factor $k$ compared to their peers in the group: it is a measure of the $k$th  factor exposure heterogeneity. A value close to one indicates a large beta heterogeneity in the peer group for factor $k$, leaving more room for investment managers to differentiate themselves from other managers regarding their factor-$k$'s exposure.

Steps 1--3 are repeated every month starting from January 2009, yielding a time series of factor exposure heterogeneity measures per factor. The ending date depends on the horizon $h$ considered: September 2020 for $h=3$, June 2020 for $h=6$, and December 2019 for $h=12$ months. Note that, by construction, the monthly ratios are autocorrelated, as they are based on overlapping daily returns (\eg, in the $h=12$ months horizon, 11 months of daily returns are overlapping between two consecutive months in the time series of the heterogeneity factor exposures).

\section{Results}
\label{sec:ratios}

\noindent
Before discussing the heterogeneity of the factor exposures, we look at the level of the exposures for green and brown stocks. Results are reported in Table~\ref{tab:exposure}. Panel A displays results for the three-month estimation window. First, as expected, we find that both green and brown stocks are primarily exposed to the market factor. Green firms are slightly more exposed with an average (computed over the estimation windows) cross-sectional median beta of 1.07, compared to brown firms at 0.92. Second, we find that green firms are positively exposed to the value factor (median beta at 0.12). Third, brown firms are positively exposed to the investment factor and negatively to the momentum factor. These findings are in line with \citet[Table 7]{ArdiaEtAl202x} and hold for the six- and twelve-month estimation windows (Panels B and C).

\insertfloat{Table~\ref{tab:exposure}}

Next, we turn to the analysis of the factor exposure heterogeneity. Table~\ref{tab:ratios} reports the average, standard deviation, maximum, and minimum values (computed over the estimation windows) of the heterogeneity measure $1-\pi_k^0$ for the six factors. We discuss results for the three-month horizon as conclusions hold for longer time horizons. 

First, the average factor exposure heterogeneity varies between 26\% and 49\%. The largest values are observed for the value factor in green stocks (49\%) and the momentum factor in brown stocks (49\%). The lowest value is observed for the size factor in green stocks (26\%). Second, we notice that, for all factors but the value factor, the heterogeneity is lower for green stocks than for brown stocks. Hence, managers shifting their investments to green stocks have less room to differentiate themselves regarding their factor exposures. Third, when looking at the variability of the exposure heterogeneity, as measured by the standard deviation or the minimum-maximum range, we see that variations in exposure heterogeneity are more important for brown stocks. Thus, the opportunity set of factor exposures varies more for brown stocks.

\insertfloat{Table~\ref{tab:ratios}}

\begin{figure}[H]
\caption{\textbf{Factor Exposure Heterogeneity Over Time}\\
The following graphs present the evolution of the six factor exposure heterogeneity $1-\pi_k^0$ over time for the brown stocks' universes. The six-factor model is estimated with twelve-month daily returns. 
The dashed line displays the time trend.}
\label{fig:ratios}
\centering
\includegraphics[width=8cm,,height=7cm]{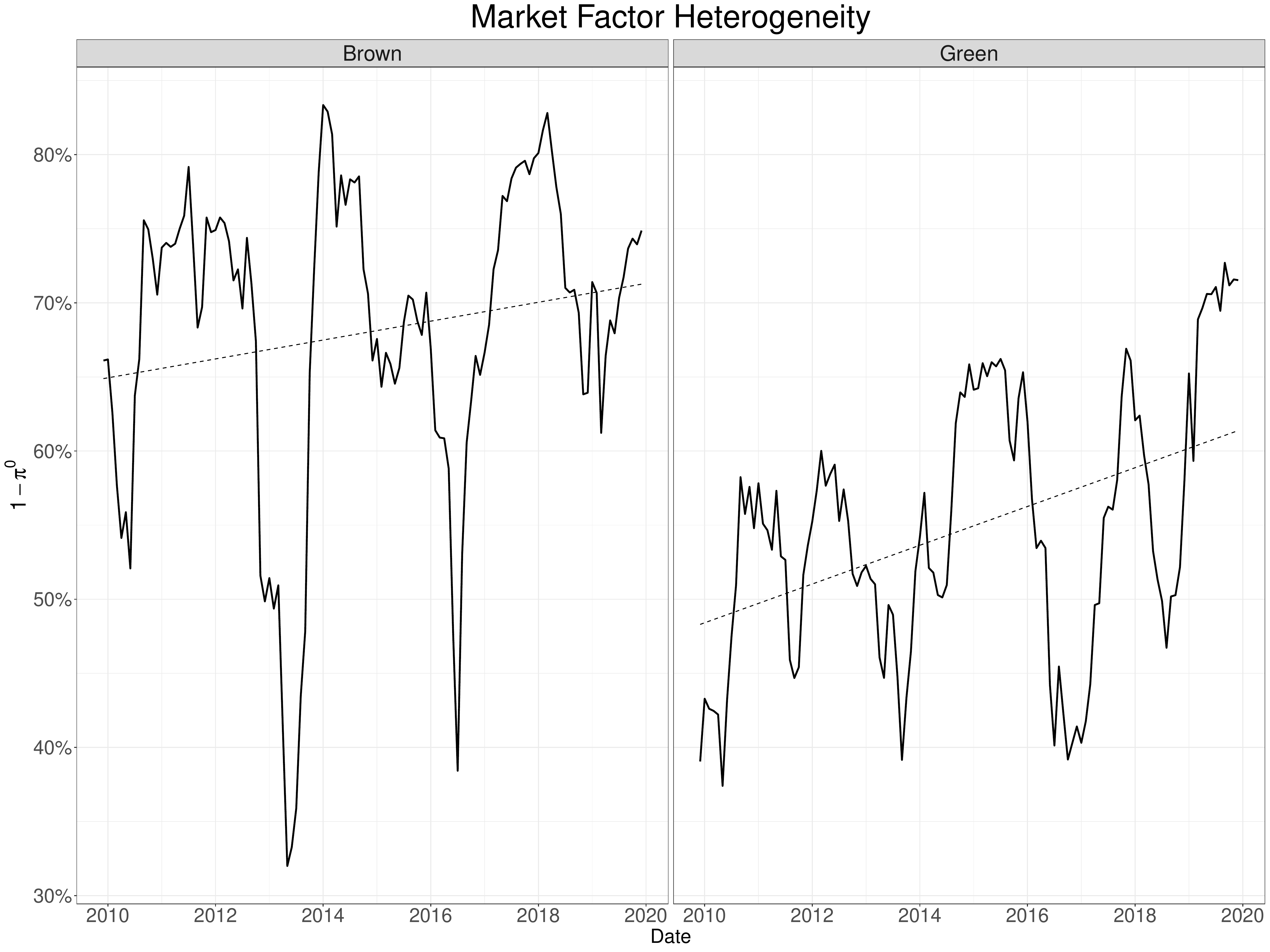}
\includegraphics[width=8cm,,height=7cm]{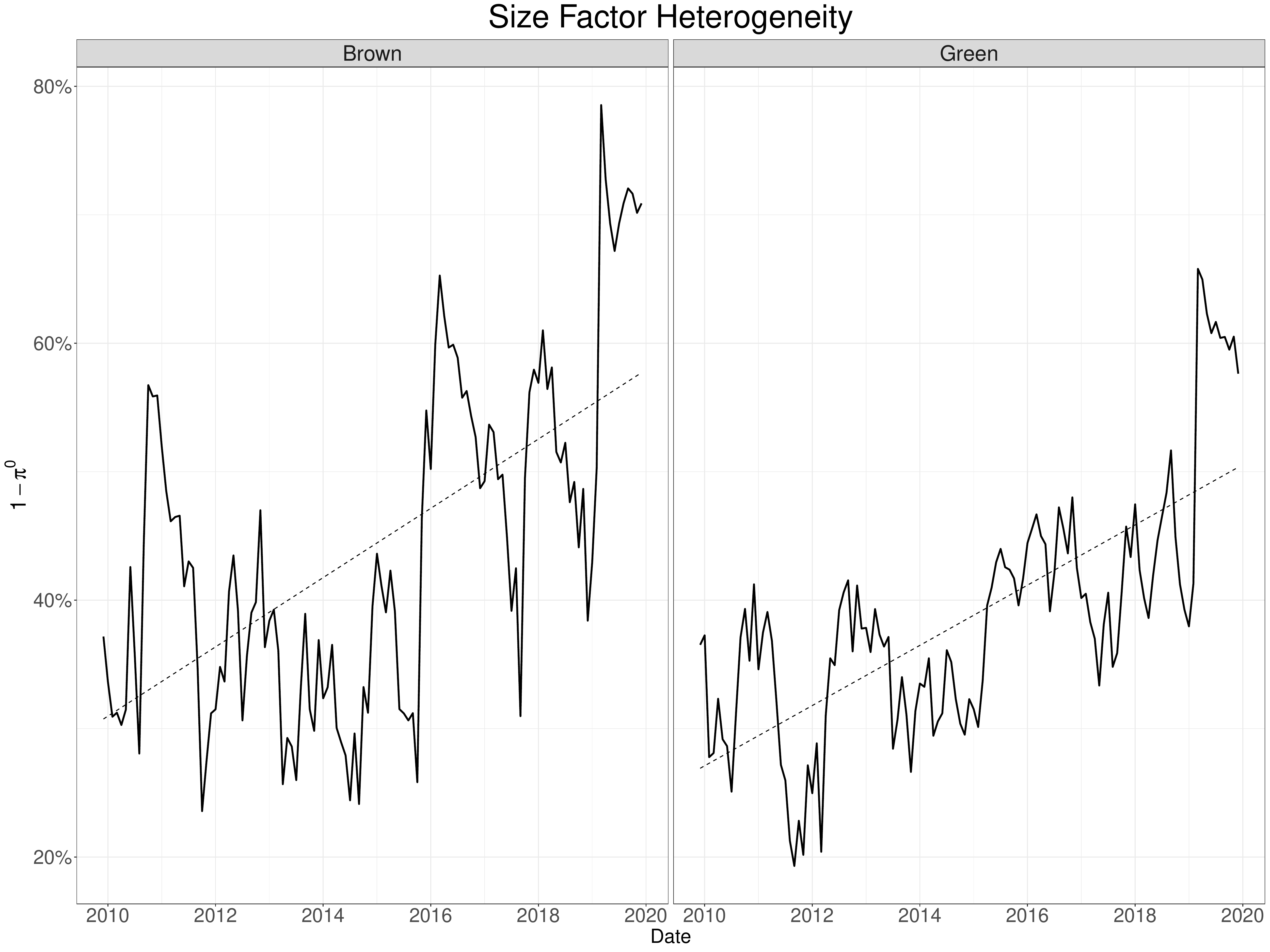}\\
\includegraphics[width=8cm,,height=7cm]{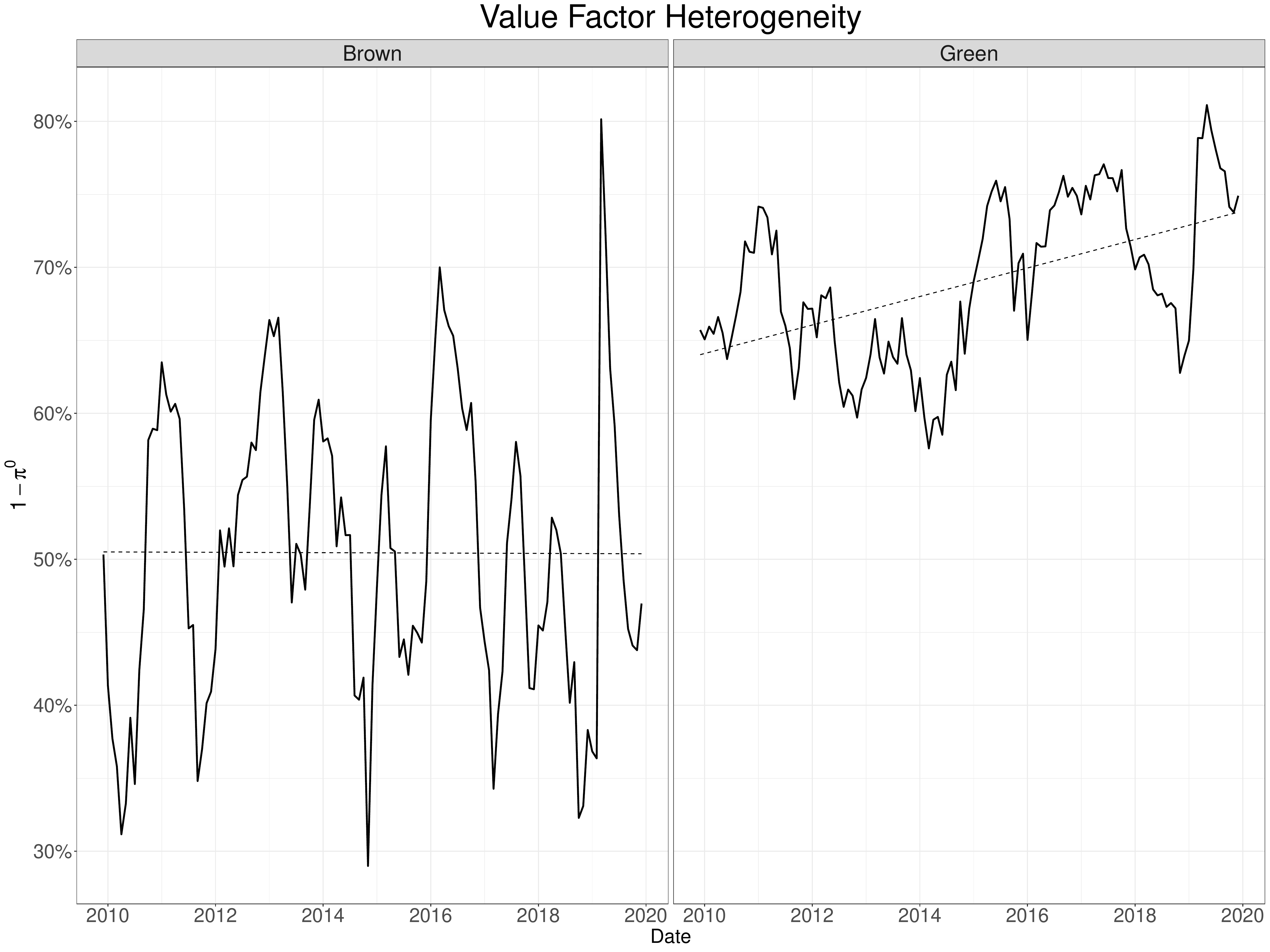}
\includegraphics[width=8cm,,height=7cm]{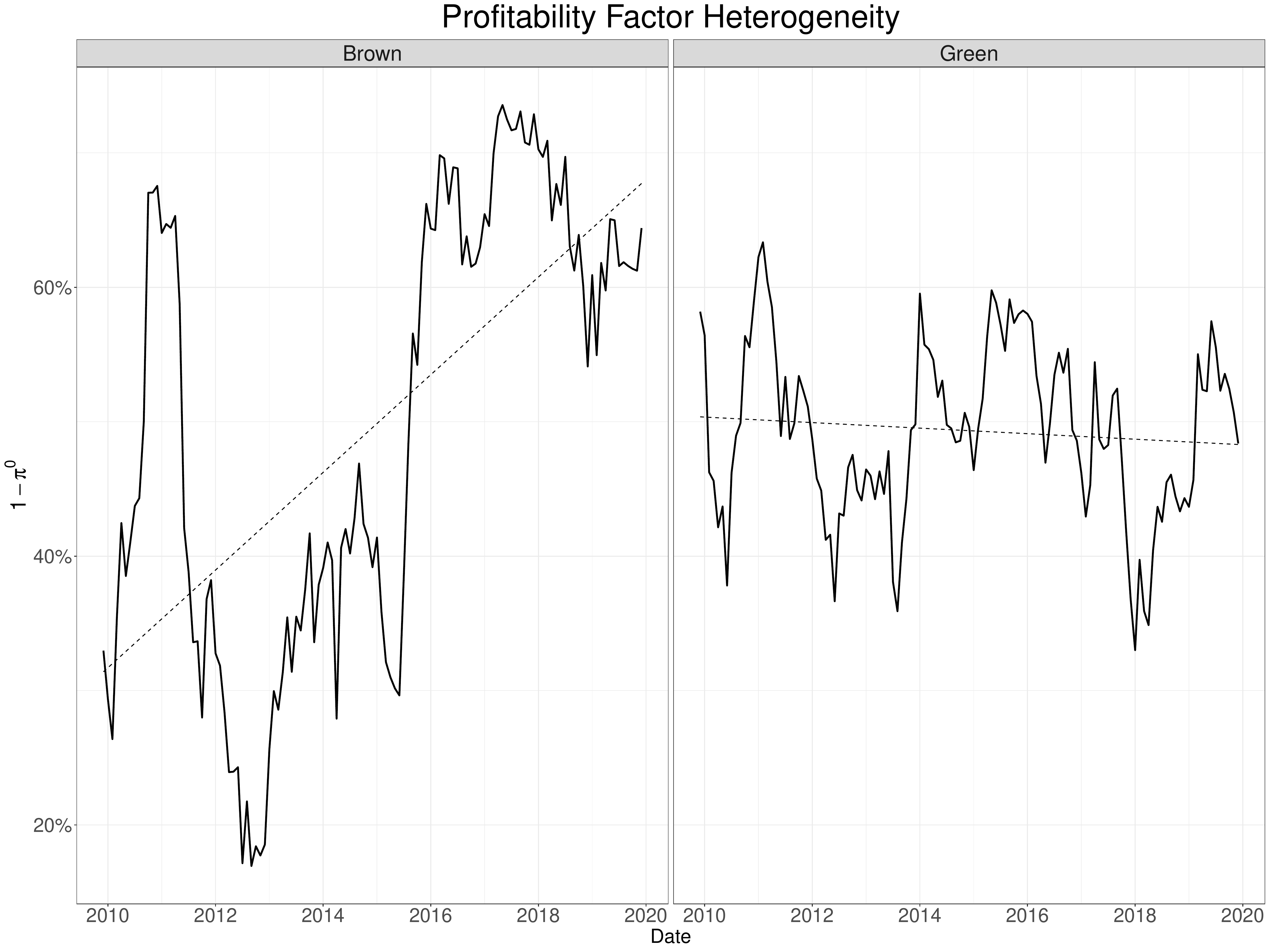}\\
\includegraphics[width=8cm,,height=7cm]{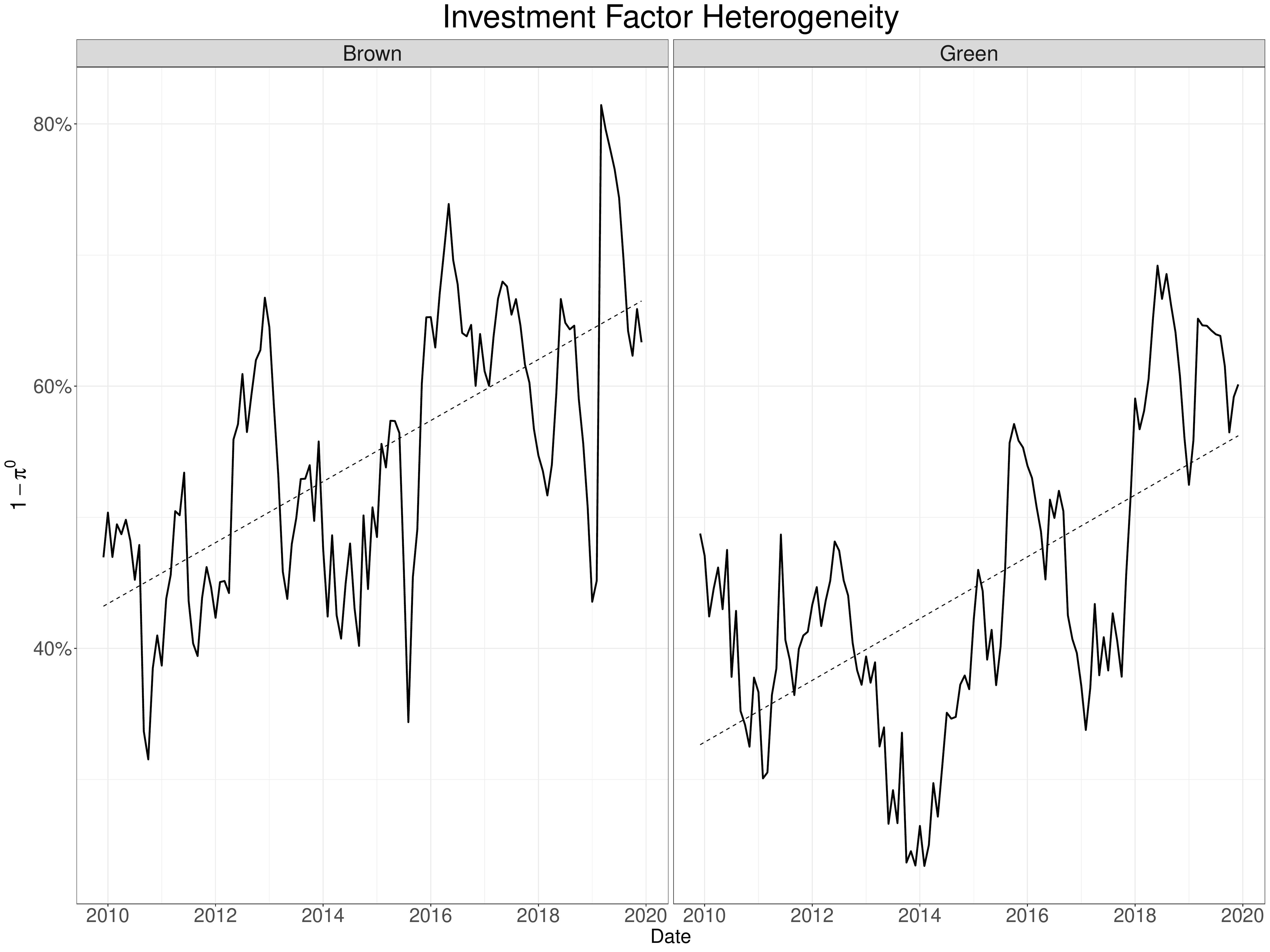}
\includegraphics[width=8cm,,height=7cm]{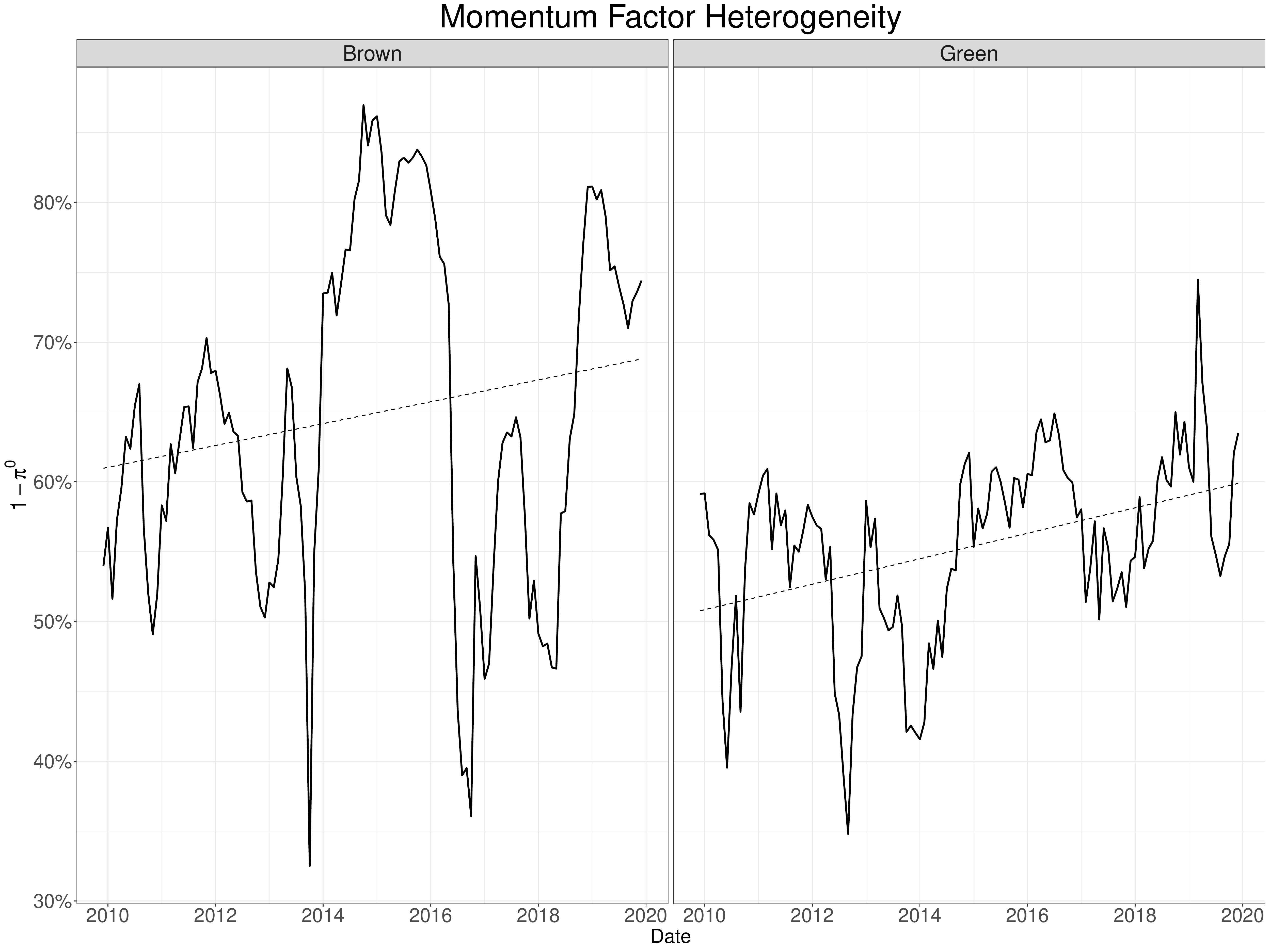}
\end{figure}

In Figure~\ref{fig:ratios}, we display the time series of the factor exposure heterogeneity of the six factors for brown and green stocks. We see a positive trend in heterogeneity for the market, size, investment, and momentum factors in both universes. The trend is also positive for the value factor in green stocks and the profitability factor in brown stocks. We can conclude that investment managers nowadays have more opportunities to differentiate themselves from others in their exposures to most equity factors, while the opportunity set has become larger for value exposures in green stocks and profitability exposures in brown stocks.

\insertfloat{Figure~\ref{fig:ratios}}

\section{Conclusion}

\noindent
Investment professionals have been facing increased constraints on their investment universes, especially since the introduction of ESG considerations into their mandate. These limitations can translate into (i) reduced alpha opportunities and/or (ii) limited exposures to traditional factors. \citet{ArdiaEtAl2022} found a significant decrease in alpha heterogeneity for green stocks in recent years ---a \emph{bad} news for managers shifting their investments to green stocks as they cannot differentiate themselves from their peers in terms of alpha (\ie, unconditional performance). In the present note, we looked at the beta component and found that beta heterogeneity is lower for green stocks than brown stocks for most traditional equity factors ---again, a \emph{bad} news for these managers, as they have less room to differentiate in terms of beta either. It is especially true for managers deploying smart beta/factor investing solutions, for which performance comes from the beta component. However, our analyses also reveal two \emph{good} news. First, the beta heterogeneity is high for the value factor ---managers can, therefore, tilt their portfolios of green stocks with high- and low-exposure stocks to the value factor should they want to differentiate from others. Second, for most factors, heterogeneity has increased in the recent period.

Overall, our analyses in \citet{ArdiaEtAl2022} and this note highlight differences between green and brown universes regarding 
alpha and beta heterogeneities. In future
work, we intend to investigate why we have differences between these groups of stocks and the potential implications for investors.\footnote{We are grateful to a reviewer for suggesting this avenue of research.}

\newpage
\begin{table}[H]
\centering
\singlespacing
\caption{\textbf{Summary of the Factor Exposures}\\
\footnotesize This table reports the summary statistics of the stock returns factor exposures in a linear model with the five factors of \citet{FamaFrench2015} augmented with the momentum factor from \citet{Carhart1997}. Specifically, the six-factor model contains the market (MKT), the size (small-minus-big or SMB), the value (high-minus-low or HML), the profitability (robust-minus-weak or RMW), the investment (conservative-minus-aggressive or CMA), and the momentum (MOM) factors. For each factor, we report the average value (over the rolling estimation windows) of the cross-sectional (across stocks) percentiles (10th, 50th, and 90th percentile) of the estimated betas and the average (over the rolling estimation windows) proportion of stocks for which the factor exposure is significantly different from zero ($\neq  0$). Significance is at the 5\% level and is based on heteroscedasticity and autocorrelation robust standard error estimators \citep{Andrews1991,AndrewsMonahan1992}. Linear regression is estimated with the three-month daily returns following the evaluation month. Brown (Green) refers to the peer group of firms in the top 75th percentile (bottom 25th percentile) of the GHG emissions at the evaluation date. Estimation windows are estimated on a rolling-window basis for a period ranging from January 2009 to September 2020.}
\begin{tabular}{lcccccccc}
\toprule
&\multicolumn{8}{c}{Panel A: Three-Month Time Horizon} \\
&\multicolumn{4}{c}{Brown} 
&\multicolumn{4}{c}{Green} \\
\cmidrule(lr){2-5}\cmidrule(lr){6-9}
&\multicolumn{3}{c}{Percentile} &&\multicolumn{3}{c}{Percentile}\\
\cmidrule(lr){2-4}\cmidrule(lr){6-8}
 & 10th & 50th & 90th & $\neq  0$ & 10th & 50th & 90th & $\neq  0$\\ 
  \midrule
  MKT & 0.43 & 0.92 & 1.50 & 0.88 & 0.62 & 1.07 & 1.57 & 0.93\\ 
  SMB & -0.71 & -0.09 & 0.61 & 0.15 & -0.58 & -0.01 & 0.62 & 0.11 \\ 
  HML & -1.01 & -0.03 & 1.04 & 0.13 & -1.01 & 0.12 & 1.39 & 0.30 \\ 
  CMA & -1.50 & 0.35 & 2.19 & 0.21 & -1.44 & -0.01 & 1.37 & 0.15 \\ 
  RMW & -1.38 & 0.09 & 1.22 & 0.19 & -1.27 & -0.08 & 1.02 & 0.15 \\ 
  MOM & -1.31 & -0.11 & 0.72 & 0.31 & -0.79 & -0.02 & 0.67 & 0.21 \\ 
  \midrule
&\multicolumn{8}{c}{Panel B: Six-Month Time Horizon} \\
&\multicolumn{4}{c}{Brown} 
&\multicolumn{4}{c}{Green} \\
\cmidrule(lr){2-5}\cmidrule(lr){6-9}
&\multicolumn{3}{c}{Percentile} &&\multicolumn{3}{c}{Percentile}\\
\cmidrule(lr){2-4}\cmidrule(lr){6-8}
 & 10th & 50th & 90th & $\neq  0$ & 10th & 50th & 90th & $\neq  0$\\ 
  \midrule
  MKT & 0.49 & 0.93 & 1.46 & 0.96 & 0.67 & 1.07 & 1.52 & 0.99\\ 
  SMB & -0.64 & -0.09 & 0.55 & 0.18 & -0.50 & -0.01 & 0.53 & 0.14 \\ 
  HML & -0.95 & -0.04 & 0.95 & 0.19 & -0.89 & 0.11 & 1.33 & 0.38 \\ 
  CMA & -1.35 & 0.38 & 2.10 & 0.29 & -1.29 & -0.02 & 1.21 & 0.19 \\ 
  RMW & -1.29 & 0.09 & 1.07 & 0.24 & -1.13 & -0.08 & 0.89 & 0.20 \\ 
  MOM & -1.19 & -0.11 & 0.64 & 0.38 & -0.69 & -0.02 & 0.56 & 0.26 \\ 
  \midrule
&\multicolumn{8}{c}{Panel C: Twelve-Month Time Horizon} \\
&\multicolumn{4}{c}{Brown} 
&\multicolumn{4}{c}{Green} \\
\cmidrule(lr){2-5}\cmidrule(lr){6-9}
&\multicolumn{3}{c}{Percentile} &&\multicolumn{3}{c}{Percentile}\\
\cmidrule(lr){2-4}\cmidrule(lr){6-8}
 & 10th & 50th & 90th & $\neq  0$ & 10th & 50th & 90th & $\neq  0$\\ 
  \midrule
  MKT & 0.47 & 0.92 & 1.48 & 0.98 & 0.64 & 1.07 & 1.54 & 0.99\\ 
  SMB & -0.67 & -0.10 & 0.53 & 0.26 & -0.54 & 0.00 & 0.59 & 0.17 \\ 
  HML & -0.98 & -0.05 & 0.95 & 0.29 & -0.95 & 0.12 & 1.38 & 0.46 \\ 
  CMA & -1.38 & 0.36 & 2.10 & 0.42 & -1.34 & -0.01 & 1.32 & 0.24 \\ 
  RMW & -1.34 & 0.11 & 1.08 & 0.32 & -1.16 & -0.05 & 0.97 & 0.27 \\ 
  MOM & -1.12 & -0.11 & 0.63 & 0.44 & -0.69 & 0.01 & 0.58 & 0.33 \\ 
\bottomrule
\end{tabular}
\label{tab:exposure}
\end{table}

\newpage
\begin{table}[H]
\singlespacing
\caption{\textbf{Summary the Factor Exposure Heterogeneity}\\
\footnotesize This table reports the summary statistics of the factor exposure heterogeneity $1 - \pi_k^0$. The different panels presents the results for each factor in our six-factor model. Three-month refers to heterogeneity computed with the three-month daily returns following the evaluation month. Six-month and one-year refer to six months and twelve months of daily returns, respectively. Brown (Green) refers to the peer group of firms in the top 75th percentile (bottom 25th percentile) of the GHG emissions at the evaluation date. Setting with $h = 3$ months spans January 2014 to September 2020 (81 evaluation dates),$h = 6$ months spans January 2014 to June 2020 (79 evaluation dates), and $h = 12$ months spans January 2014 to December 2019 (72 evaluation dates).}
\centering
\vspace{-0.5cm}
\scalebox{0.82}{
\begin{tabular}{lrrrrrr}
\toprule
\multicolumn{7}{l}{Panel A:  Market Factor (MKT)} \\ 
&\multicolumn{2}{c}{$h = 3$ months} 
&\multicolumn{2}{c}{$h = 6$ months} 
&\multicolumn{2}{c}{$h = 12$ months} \\
\cmidrule(lr){2-3}\cmidrule(lr){4-5}\cmidrule(lr){6-7}
& Brown  & Green  & Brown  & Green  & Brown  & Green  \\ 
\midrule
Average & 0.43 & 0.35 & 0.55 & 0.43 & 0.68 & 0.55 \\ 
Standard  deviation   & 0.14 &  0.08 &  0.13 &  0.10 &  0.11 &  0.09 \\ 
Minimum &  0.16 &  0.18 &  0.21 & 0.25 &  0.32 & 0.37 \\ 
Maximum & 0.74 & 0.55 & 0.79 & 0.65 & 0.83 & 0.73 \\ 
\midrule
\multicolumn{7}{l}{Panel B:  Size Factor (SMB)} \\ 
&\multicolumn{2}{c}{$h = 3$ months} 
&\multicolumn{2}{c}{$h = 6$ months} 
&\multicolumn{2}{c}{$h = 12$ months} \\
\cmidrule(lr){2-3}\cmidrule(lr){4-5}\cmidrule(lr){6-7}
& Brown  & Green  & Brown  & Green  & Brown  & Green  \\ 
\midrule
Average & 0.29 & 0.26 & 0.35 & 0.32 & 0.44 & 0.39 \\ 
Standard  deviation   & 0.09 &  0.07 &  0.11 &  0.08 &  0.13 &  0.09 \\ 
Minimum &  0.13 &  0.14 &  0.17 & 0.17 &  0.24 & 0.19 \\ 
Maximum & 0.63 & 0.46 & 0.74 & 0.62 & 0.79 & 0.66 \\ 
\midrule
\multicolumn{7}{l}{Panel C:  Value Factor (HML)} \\ 
&\multicolumn{2}{c}{$h = 3$ months} 
&\multicolumn{2}{c}{$h = 6$ months} 
&\multicolumn{2}{c}{$h = 12$ months} \\
\cmidrule(lr){2-3}\cmidrule(lr){4-5}\cmidrule(lr){6-7}
& Brown  & Green  & Brown  & Green  & Brown  & Green  \\ 
\midrule
Average & 0.32 & 0.49 & 0.41 & 0.60 & 0.50 & 0.69 \\ 
Standard  deviation   & 0.10 &  0.09 &  0.11 &  0.08 &  0.10 &  0.06 \\ 
Minimum &  0.14 &  0.29 &  0.22 & 0.35 &  0.29 & 0.58 \\ 
Maximum & 0.68 & 0.67 & 0.75 & 0.73 & 0.80 & 0.81 \\ 
\midrule
\multicolumn{7}{l}{Panel D:  Investment Factor (CMA)} \\ 
&\multicolumn{2}{c}{$h = 3$ months} 
&\multicolumn{2}{c}{$h = 6$ months} 
&\multicolumn{2}{c}{$h = 12$ months} \\
\cmidrule(lr){2-3}\cmidrule(lr){4-5}\cmidrule(lr){6-7}
& Brown  & Green  & Brown  & Green  & Brown  & Green  \\ 
\midrule
Average & 0.36 & 0.30 & 0.45 & 0.37 & 0.55 & 0.44 \\ 
Standard  deviation   & 0.12 &  0.08 &  0.12 &  0.09 &  0.10 &  0.11 \\ 
Minimum &  0.16 &  0.12 &  0.17 & 0.19 &  0.32 & 0.23 \\ 
Maximum & 0.72 & 0.55 & 0.78 & 0.63 & 0.81 & 0.69 \\ 
\midrule
\multicolumn{7}{l}{Panel E:  Profitability Factor (RMW)} \\ 
&\multicolumn{2}{c}{$h = 3$ months} 
&\multicolumn{2}{c}{$h = 6$ months} 
&\multicolumn{2}{c}{$h = 12$ months} \\
\cmidrule(lr){2-3}\cmidrule(lr){4-5}\cmidrule(lr){6-7}
& Brown  & Green  & Brown  & Green  & Brown  & Green  \\ 
\midrule
Average & 0.35 & 0.32 & 0.41 & 0.39 & 0.50 & 0.49 \\ 
Standard  deviation   & 0.12 &  0.07 &  0.14 &  0.07 &  0.17 &  0.06 \\ 
Minimum &  0.15 &  0.17 &  0.15 & 0.23 &  0.17 & 0.33 \\ 
Maximum & 0.65 & 0.53 & 0.71 & 0.57 & 0.74 & 0.63 \\ 
\midrule
\multicolumn{7}{l}{Panel F:  Momentum Factor (MOM)} \\ 
&\multicolumn{2}{c}{$h = 3$ months} 
&\multicolumn{2}{c}{$h = 6$ months} 
&\multicolumn{2}{c}{$h = 12$ months} \\
\cmidrule(lr){2-3}\cmidrule(lr){4-5}\cmidrule(lr){6-7}
& Brown  & Green  & Brown  & Green  & Brown  & Green  \\ 
\midrule
Average & 0.49 & 0.37 & 0.58 & 0.46 & 0.65 & 0.55 \\ 
Standard  deviation   & 0.14 &  0.09 &  0.13 &  0.09 &  0.12 &  0.07 \\ 
Minimum &  0.19 &  0.17 &  0.20 & 0.22 &  0.33 & 0.35 \\ 
Maximum & 0.75 & 0.64 & 0.82 & 0.66 & 0.87 & 0.74 \\ 
\bottomrule
\end{tabular}}
\label{tab:ratios}
\end{table}

\newpage

\end{document}